\documentclass[preprint,proceedings]{rmaa}

\SetYear{2002}
\SetConfTitle{Galaxy Evolution: Theory and Observations}

\title{The blue colors of Gamma-ray burst host galaxies}

\author{
  E. Le Floc'h\altaffilmark{1}, P.-A. Duc\altaffilmark{1} and 
I.F. Mirabel\altaffilmark{1,2}}

\altaffiltext{1}{CEA/DSM/DAPNIA Service d'Astrophysique, F-91191
Gif-sur-Yvette,  France.}
\altaffiltext{2}{Instituto de Astronom\'\i a y F\'\i sica del 
Espacio, cc 67, suc 28. 1428 Buenos Aires, Argentina.}

\suppressfulladdresses

\listofauthors{ E. Le Floc'h, P.-A. Duc and I.F. Mirabel}
\indexauthor{Le Floc'h, E., Duc, P.-A., Mirabel, I.F.}

\addkeyword{galaxies: starburst}
\addkeyword{galaxies: evolution}
\addkeyword{gamma-rays: bursts}

\begin{document}
\maketitle 

\boldabstract{There is now an increasing number of evidence 
supporting the idea that
the cosmic Gamma-ray Bursts (GRBs) originate from the collapse
of massive stars in distant star-forming galaxies. 
Because GRBs are
likely detectable up to very high redshift, and because the gamma-rays
are not attenuated by intervening columns of gas and dust, these
phenomena thus offer a unique perspective to probe the star formation
in the early Universe independently of the biases associated with dust
extinction.
}

In a first step of a long-term study to characterize the physical
properties of the starbursts pinpointed by GRBs, we report here on the
R--K colors of GRB host galaxies. Such colors, in principal, should
provide indications about the fraction of star formation occuring in
blue galaxies as opposed to that taking place in highly reddened
sources.  In Figure\,1 we compare the R--K colors of GRB hosts with
those of field galaxies from the HDF (Fernandez-Soto et al. 1999). The
K magnitudes of GRB hosts were partly derived from our near-infrared
(NIR) observations at the VLT, whereas the remaining data points and
the R magnitudes were gathered from the literature.  Using the optical
and NIR local templates of Coleman et al. (1980) and Mannucci et
al. (2001), we have also overplotted the colors that would exhibit
such templates if they were shifted to higher z assuming no
evolution. One can clearly see that GRB hosts display rather blue
colors, typical of the population of blue faint sources at high
z. They moreover appear significantly bluer than the local irregular
galaxies, which reveals an even stronger
contribution of the
UV-continuum from star-forming regions.

On one hand, this result seems to be consistent with the blue colors
characteristic of the active star-forming HII regions that dominate
the optical emission of starburst galaxies at high redshift. On the
other hand, several authors have recently argued that a significant
fraction of the star formation history may have occured in reddened
sources enshrouded in dusty environments, such as those found among
the Extremely Red Objects (EROs, R--K $\geq$ 5). Thanks to their
dust-penetrating power, GRBs should be capable to trace such a hidden
star-forming activity.  So far indeed, GRBs have enabled the discovery
of two submillimeter dusty galaxies, yet with blue optical
counterparts, but surprisingly no ERO has been observed
among the GRB hosts (see Figure\,1).  A plausible explanation may
simply originate from the very small number of sources in the current
sample of GRB host galaxies. Nonetheless, if this tendancy was to be
confirmed with a larger number of sources, key results could be
deduced and -- unless a careful revision of the GRB starburst
selection and the possible effects of metallicity in the burst
environments is to be done -- severe constraints relative to the
contribution of reddened starbursts at high z may indeed be obtained.

The advent of future
instruments dedicated to GRBs and their afterglows
will provide a better statistics 
to investigate further this issue.

\begin{figure}[!t]
  \includegraphics[width=\columnwidth]{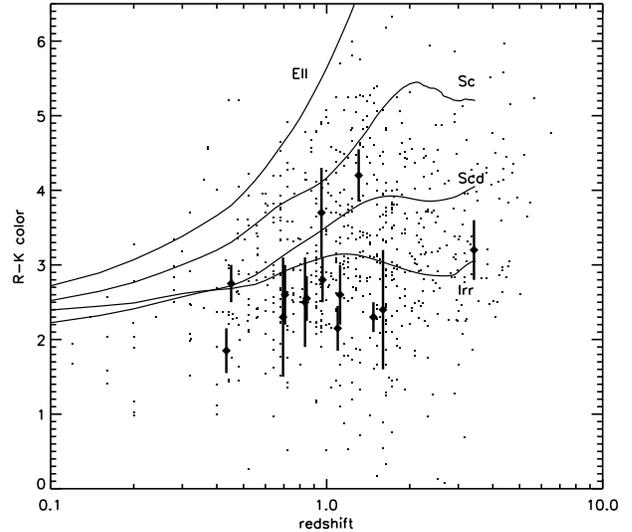}
  \caption{The R--K colors of GRB host galaxies (filled
diamonds), compared with those of field sources from
the HDF (dots). Solid lines represent the expected colors
of local templates shifted to higher z.

}
  \label{fig:color}
\end{figure}


\end{document}